\documentclass{article}
\usepackage{spconf,amsmath,amssymb}
\usepackage{graphicx}
\usepackage{algorithm, algorithmic}
\usepackage{textcomp}
\usepackage{booktabs}
\usepackage{subcaption}
\usepackage{siunitx}
\usepackage{float}
\usepackage{hyperref}
\usepackage{xcolor}

\newcommand{\mat}[1]{\ensuremath{{\mathbf{\MakeUppercase{#1}}}}}
\newcommand{\transpose}[1]{\ensuremath{{#1}^{\textsc{t}}}}
\newcommand{\trace}[1]{\ensuremath{\text{Tr}\left(#1\right)}}

\DeclareMathOperator*{\argmax}{arg\,max}

\title{Efficient Solutions for Mitigating Initialization Bias in Unsupervised Self-Adaptive Auditory Attention Decoding}
\name{Yuanyuan Yao$^1$, Simon Geirnaert$^{1,2}$, Tinne Tuytelaars$^{3}$, Alexander Bertrand$^{1}$\thanks{This research is funded by the Research Foundation - Flanders (FWO) project No G081722N, junior postdoctoral fellowship fundamental research of the FWO (for S. Geirnaert, No. 1242524N), the European Research Council (ERC) under the European Union's Horizon 2020 research and innovation program (grant agreement No 101138304), Internal Funds KU Leuven (projects IDN/23/006, C14/25/108, and C3/25/107), and the Flemish Government (AI Research Program). Views and opinions expressed are however those of the author(s) only and do not necessarily reflect those of the European Union or the granting authorities. Neither the European Union nor the granting authorities can be held responsible for them.\\All authors are also affiliated with Leuven.AI - KU Leuven institute for AI, Belgium.}}
\address{$^1$KU Leuven, Department of Electrical Engineering (ESAT), \\STADIUS Center for Dynamical Systems, Signal Processing and Data Analytics, Belgium\\
$^2$KU Leuven, Department of Neurosciences, Research Group ExpORL, Belgium\\
$^3$KU Leuven, Department of Electrical Engineering (ESAT), PSI, Belgium\\
}
\begin{document}
%\ninept
%
\maketitle
\begin{abstract}
Decoding the attended speaker in a multi-speaker environment from electroencephalography (EEG) has attracted growing interest in recent years, with neuro-steered hearing devices as a driver application. Current approaches typically rely on ground-truth labels of the attended speaker during training, necessitating calibration sessions for each user and each EEG set-up to achieve optimal performance. While unsupervised self-adaptive auditory attention decoding (AAD) for stimulus reconstruction has been developed to eliminate the need for labeled data, it suffers from an initialization bias that can compromise performance. Although an unbiased variant has been proposed to address this limitation, it introduces substantial computational complexity that scales with data size. This paper presents three computationally efficient alternatives that achieve comparable performance, but with a significantly lower and constant computational cost. The code for the proposed algorithms is available at \url{https://github.com/YYao-42/Unsupervised_AAD}.
\end{abstract}
\section{Introduction}
\label{sec:intro}

\par Auditory attention decoding (AAD) aims to identify the attended speaker in complex multi-speaker auditory environments from brain signals such as electroencephalography (EEG). This can be useful for neuro-steered hearing aids, where the attended speaker's volume can be enhanced while suppressing unattended speakers \cite{geirnaert2021electroencephalography, pan2024neuroheed}. A common approach trains a linear neural decoder to reconstruct features of the attended speech from EEG, and during testing, the attended speaker is identified as the candidate most correlated with the reconstruction \cite{o2015attentional, wong2018comparison, ciccarelli2019comparison, alickovic2019tutorial}. However, training or fine-tuning a neural decoder requires labeled data indicating the attended speaker at each time point, which in practice necessitates a calibration session where the user follows attention instructions. Such sessions are a nuisance, in particular if they have to be repeated frequently.

\par To address this, Geirnaert et al. \cite{Geirnaert2021Unsupervised} proposed unsupervised AAD based on stimulus reconstruction, beginning with training a decoder with random attention labels. Even when using random labels in this initial training phase, it was shown that the resulting decoder can achieve above-chance decoding performance. Each subsequent iteration updates the model using newly predicted labels, which include a higher proportion of true attended labels. This yields a higher-quality decoder that produces more accurate predictions in the next iteration. This bootstrapping effect continues until convergence.

\par However, this approach suffers from initialization bias, as training and testing on the same data across iterations causes the model to favor initial (possibly wrong) predictions \cite{Heintz2023Unbiased}. Heintz et al. addressed this by implementing leave-one-out cross-validation within each iteration \cite{Heintz2023Unbiased}: training on $K-1$ segments and predicting the remaining segment, repeated for all $K$ segments. While this cross-validated version improves performance, particularly with limited training data, it requires training the model $K$ times per iteration, creating substantial computational overhead.

\par In this paper, we propose unsupervised training methods based on canonical
correlation analysis (CCA) that are inherently robust to AAD label noise in such self-adaptive iterations. Our approach achieves comparable performance to the cross-validated version from \cite{Heintz2023Unbiased} but requires only one model training per iteration instead of $K$. This substantial reduction in computation time makes it particularly well-suited for real-time or time-adaptive implementations \cite{geirnaert2022time}.

\section{Methods}
\label{sec:methods}

\par We consider a 2-speaker scenario without loss of generality as all methods can be straightforwardly extended to an $N$-speaker setting. Let $\mathbf{S}_1, \mathbf{S}_2 \in \mathbb{R}^{T \times D_s}$ denote the speech features, such as speech envelopes, of the two candidate speakers, where $T$ is the number of samples and $D_s$ is the feature dimension. Given EEG signals $\mathbf{X} \in \mathbb{R}^{T \times D_x}$ with $D_x$ channels, the goal is to determine which of $\mathbf{S}_1$ or $\mathbf{S}_2$ corresponds to the attended speaker's features (denoted $\mathbf{S}_a$) and which to the unattended speaker's features (denoted $\mathbf{S}_u$). In practice, decoding is performed on a segment basis to obtain time-resolved estimates of attention: a long recording is divided into $K$ segments, yielding $\{\mathbf{X}_k\}_{k=1}^K$, $\{\mathbf{S}_{1k}\}_{k=1}^K$, and $\{\mathbf{S}_{2k}\}_{k=1}^K$, and the attended speaker is identified per segment. For simplicity, we assume the signals are centered, i.e., $\mathbb{E}[{\mathbf X}_k] = \mathbb{E}[\mathbf S_{1k}] = \mathbb{E}[\mathbf S_{2k}] = \mathbf 0$, where $\mathbb{E}[\cdot]$ denotes the expectation operator.

\subsection{Baseline: Single-Encoder Version}
\label{sec:method_se}

\par The original algorithm in \cite{Geirnaert2021Unsupervised} is based on a backward model that reconstructs the attended speaker features from EEG signals using linear regression. Here we present a more general CCA-based variant based on \cite{de2018decoding}. In the supervised setting, CCA optimizes a decoder $\mathbf w_x\in \mathbb R^{D_x \times 1}$ (on the EEG side) and an encoder $\mathbf w_a\in \mathbb R^{D_s \times 1}$ (on the audio side) to maximize the correlation between the transformed EEG signals and the features of the attended speaker:
\begin{equation}
  \begin{aligned}
  & \underset{\mathbf{w}_x, \mathbf{w}_a}{\text{maximize}}
  & & \transpose{\mathbf w}_x \transpose{\mathbf X} \mathbf S_a\mathbf w_a \\
  & \text{subject to}
  & & \transpose{\mathbf w}_x \transpose{\mathbf X} \mathbf X \mathbf w_x=1, \\
  &&& \transpose{\mathbf w}_a \transpose{\mathbf S_a} \mathbf S_a\mathbf w_a=1.
  \end{aligned}
  \label{eq:cca}
\end{equation}
As opposed to the backward model, CCA allows exploiting both multivariate data modalities to identify a shared subspace that maximizes the correlation between them.

\par The optimized vectors $\hat{\mathbf w}_x$ and $\hat{\mathbf w}_a$ are the first canonical components. Higher-order components are obtained iteratively by solving \eqref{eq:cca} in a subspace where the transformed signals are orthogonal to those from previous iterations. In matrix form, this corresponds to:
\begin{equation}
  \begin{aligned}
  & \underset{\mathbf{W}_x, \mathbf{W}_a}{\text{maximize}}
  & & \trace{\transpose{\mat{W}}_x\mat{R}_{xa}\mat{W}_a} \\
  & \text{subject to}
  & & \transpose{\mat{W}}_x\mat{R}_{xx}\mat{W}_x = \mat{I}_Q, \\
  &&& \transpose{\mat{W}}_a\mat{R}_{aa}\mat{W}_a = \mat{I}_Q,
  \end{aligned}
  \label{eq:cca_multi}
\end{equation}
where $\mathbf W_x =[\mathbf w_{x1} \cdots \mathbf w_{xQ}]$ and $\mathbf W_a =[\mathbf w_{a1} \cdots \mathbf w_{aQ}]$ contain the first $Q$ canonical components, $\mat{R}_{xa} = \transpose{\mathbf X}\mathbf S_a$, $\mat{R}_{xx} = \transpose{\mathbf X}\mathbf X$, and $\mat{R}_{aa} = \transpose{\mathbf S_a}\mathbf S_a$. The solution to \eqref{eq:cca_multi} can be obtained by solving a generalized eigenvalue decomposition (GEVD) problem \cite{corrochano2005handbook}:
\begin{equation}
    \label{eq:cca_gevd}
    \mathbf R \hat{\mathbf W} = \mathbf D \hat{\mathbf W} \boldsymbol \Lambda,
\end{equation}
where $\boldsymbol \Lambda$ is a diagonal matrix containing the generalized eigenvalues ordered in descending order, and 
\begin{equation}
  \mathbf R = \begin{bmatrix}
    \mathbf R_{xx} & \mathbf R_{xa} \\
    \transpose{\mathbf R}_{xa} & \mathbf R_{aa}
    \end{bmatrix}, \mathbf D= \begin{bmatrix}
        \mathbf  R_{xx} & \mathbf 0 \\
        \mathbf 0 & \mathbf  R_{aa}
        \end{bmatrix}, \hat{\mathbf W} = \begin{bmatrix}
      \hat{\mathbf W}_x \\
      \hat{\mathbf W}_a 
      \end{bmatrix}. \label{eq:single_param_gevd}
\end{equation}

\par For a test pair of features $(\mathbf{S}_{1k}, \mathbf{S}_{2k})$, the attended speaker is identified by finding 
\begin{equation}
  j = \argmax_{i \in \{1,2\}} \tilde{\rho}_{ik},
  \label{eq:att_speaker}
\end{equation}
where $\tilde{\rho}_{ik}$ is the sum of canonical correlations computed using the trained decoder $\hat{\mathbf W}_x$ and encoder $\hat{\mathbf W}_a$:
\begin{equation}
  \tilde{\rho}_{ik} = \sum_{q=1}^Q \frac{\hat{\mathbf w}_{xq}^\top \mathbf{X}_k^\top \mathbf{S}_{ik} \hat{\mathbf w}_{aq}}{\sqrt{\hat{\mathbf w}_{xq}^\top \mathbf{X}_k^\top \mathbf{X}_{k} \hat{\mathbf w}_{xq}} \sqrt{\hat{\mathbf w}_{aq}^\top \mathbf{S}_{ik}^\top \mathbf{S}_{ik} \hat{\mathbf w}_{aq}}}, \quad i=1,2.
  \label{eq:single_enc_rho}
\end{equation}
$\mathbf{S}_{jk}$ is then assigned as $\mathbf{S}_{ak}$, and the other as $\mathbf{S}_{uk}$.

\par In the unsupervised setting, the true attended speaker labels are unavailable, so the statistics $\mathbf{R}_{aa}$ and $\mathbf{R}_{xa}$ cannot be computed because the per-speaker segments $\{\mathbf S_{1k}\}_{k=1}^K$ and $\{\mathbf S_{2k}\}_{k=1}^K$ cannot be mapped to attended and unattended sets $\{\mathbf S_{ak}\}_{k=1}^K$ and $\{\mathbf S_{uk}\}_{k=1}^K$. Following the self-adaptive approach of \cite{Geirnaert2021Unsupervised}, we begin by randomly assigning one of $\mathbf{S}_{1k}$ or $\mathbf{S}_{2k}$ as $\mathbf{S}_{ak}$ (and the other as $\mathbf{S}_{uk}$). Using these initial random labels, we form $\mathbf X, \mathbf S_a$, and $\mathbf S_u$ by stacking the segments over time. $\mathbf{W}_x$ and $\mathbf{W}_a$ are then estimated by solving \eqref{eq:cca_gevd}, and the labels are updated according to \eqref{eq:att_speaker}. This procedure is iterated, each time based on the previously assigned labels, until convergence. A summary of this basic version, here called the single-encoder version, is provided in Algorithm \ref{alg:single_two_enc}. However, as identified in \cite{Heintz2023Unbiased}, this approach suffers from initialization bias where the model favors assigning the same labels as in the previous iteration, making initial wrong predictions persist. To address this without expensive inner cross-validations as in \cite{Heintz2023Unbiased}, we propose three variants described in the following sections.

\subsection{Two-Encoder Version}
\label{sec:method_te}

\par For the single-encoder version, only the features of the attended speaker $\mathbf S_a$ are incorporated in the optimization. We extend this with two encoders: $\mathbf W_a$ for the attended features $\mathbf S_a$ and $\mathbf W_u$ for the unattended features $\mathbf S_u$, both sharing the same decoder $\mathbf W_x$, to jointly maximize the correlation between EEG and the features of both speakers:
\begin{equation}
  \begin{aligned}
  & \underset{\mathbf{W}_x, \mathbf{W}_a, \mathbf{W}_u}{\text{maximize}}
  & & \trace{\transpose{\mat{W}}_x\mat{R}_{xa}\mat{W}_a+\transpose{\mat{W}}_x\mat{R}_{xu}\mat{W}_u} \\
  & \text{subject to}
  & & \transpose{\mat{W}}_x\mat{R}_{xx}\mat{W}_x = \mat{I}_Q, \\
  &&& \begin{bmatrix}\transpose{{\mathbf W}_a} & \transpose{{\mathbf W}_u} \end{bmatrix}\begin{bmatrix}
    \mathbf R_{aa} & \mathbf R_{au} \\
    \transpose{\mathbf R}_{au} & \mathbf R_{uu}
    \end{bmatrix}\begin{bmatrix}{{\mathbf W}_a} \\{{\mathbf W}_u} \end{bmatrix}=\mathbf I_{Q},
  \end{aligned}
  \label{eq:cca_two_enc}
\end{equation}
where $\mat{R}_{xu} = \transpose{\mathbf X}\mathbf S_u,\mat{R}_{au} = \transpose{\mathbf S_a}\mathbf S_u$, and $\mat{R}_{uu} = \transpose{\mathbf S_u}\mathbf S_u$.

\par In the ideal case with sufficient labeled data, the two-encoder version may suffer performance loss due to its inherent reduced discriminative power as $\mathbf W_x$ is encouraged to extract EEG responses to both speakers (not only the attended one). However, in unsupervised self-adaptive settings, this approach may be more robust to label errors, as $\mathbf W_x$ will be less attracted to wrongly labeled attended speech responses (which would otherwise bias it towards producing the same wrong labels for the next iteration). This increases the chance of recovering from wrong predictions.

\par The solution to \eqref{eq:cca_two_enc} can again be obtained by solving the GEVD problem \eqref{eq:cca_gevd}, with 
\begin{equation}
\begin{aligned}
    \mathbf R =
  \begin{bmatrix}
    \mathbf R_{xx} & \mathbf R_{xa} & \mathbf R_{xu}\\
    \transpose{\mathbf R}_{xa} & \mathbf R_{aa} & \mathbf R_{au} \\
    \transpose{\mathbf R}_{xu} & \transpose{\mathbf R}_{au} & \mathbf R_{uu}
    \end{bmatrix}&, \mathbf D = \begin{bmatrix}
        \mathbf  R_{xx} & \mathbf 0 & \mathbf 0 \\
        \mathbf 0 & \mathbf  R_{aa} & \mathbf R_{au}  \\
        \mathbf 0 & \transpose{\mathbf R}_{au} & \mathbf  R_{uu}
        \end{bmatrix}, \\
        \mathbf W =& \transpose{\begin{bmatrix}
      \transpose{\hat{\mathbf W}_x} & \transpose{\hat{\mathbf W}_a} & \transpose{\hat{\mathbf W}_u} \end{bmatrix}}. \label{eq:two_param_gevd}
\end{aligned}
\end{equation}
The attended speaker is still identified using \eqref{eq:att_speaker}.
When predicting the attended segments, however, only the attended encoder is used (together with $\mathbf W_x$). A summary can be found in Algorithm \ref{alg:single_two_enc}.

\begin{algorithm}[t]
\caption{Single-/Two-Encoder Version}
\label{alg:single_two_enc}
\footnotesize
\begin{algorithmic}[1]
\STATE \textbf{Input:} EEG segments $\{{\mathbf X}_k\}_{k=1}^K$, speaker features $\{{\mathbf S_1}_k\}_{k=1}^K$ and $\{{\mathbf S_2}_k\}_{k=1}^K$, number of components $Q$
\STATE \textbf{Initialize:} For each $k$, draw $j\in\{1,2\}$ uniformly and set $\mathbf S_{ak}\!\leftarrow\!\mathbf S_{jk}$, $\mathbf S_{uk}\!\leftarrow\!\mathbf S_{(3-j)k}$
\WHILE{not converged}
  \STATE Build $\mathbf R$, $\mathbf D$ from current labels. For single-encoder, 
  \vspace{-0.5em}
  \begin{equation*}
    \mathbf R = \begin{bmatrix}
      \mathbf R_{xx} & \mathbf R_{xa} \\
      \transpose{\mathbf R}_{xa} & \mathbf R_{aa}
      \end{bmatrix}, \mathbf D= \begin{bmatrix}
          \mathbf  R_{xx} & \mathbf 0 \\
          \mathbf 0 & \mathbf  R_{aa}
          \end{bmatrix}.
  \end{equation*}
  \vspace{-0.3em}
  For two-encoder,
  \vspace{-0.5em}
  \begin{equation*}
      \mathbf R =
  \begin{bmatrix}
    \mathbf R_{xx} & \mathbf R_{xa} & \mathbf R_{xu}\\
    \transpose{\mathbf R}_{xa} & \mathbf R_{aa} & \mathbf R_{au} \\
    \transpose{\mathbf R}_{xu} & \transpose{\mathbf R}_{au} & \mathbf R_{uu}
    \end{bmatrix}, \mathbf D = \begin{bmatrix}
        \mathbf  R_{xx} & \mathbf 0 & \mathbf 0 \\
        \mathbf 0 & \mathbf  R_{aa} & \mathbf R_{au}  \\
        \mathbf 0 & \transpose{\mathbf R}_{au} & \mathbf  R_{uu}
        \end{bmatrix}.
  \end{equation*}
  \vspace{-0.25em}
  \STATE Solve $\mathbf R \hat{\mathbf W} = \mathbf D \hat{\mathbf W}\boldsymbol\Lambda$ to obtain $\hat{\mathbf W}$ (partitioned as $\hat{\mathbf W}_x,\hat{\mathbf W}_a$ for single-encoder; $\hat{\mathbf W}_x,\hat{\mathbf W}_a,\hat{\mathbf W}_u$ for two-encoder).
  \STATE For each $k$, compute $\tilde\rho_{1k}$ and $\tilde\rho_{2k}$ using \eqref{eq:single_enc_rho}, and set $
    j= \arg\max_{i\in\{1,2\}} \tilde\rho_{ik},\quad
    \mathbf S_{ak}\!\leftarrow\!\mathbf S_{jk},\quad
    \mathbf S_{uk}\!\leftarrow\!\mathbf S_{(3-j)k}.$
\ENDWHILE
\STATE \textbf{Output:} $\hat{\mathbf W}$
\end{algorithmic}
\end{algorithm}

\begin{algorithm}[t]
\caption{Soft version}
\label{alg:soft_enc}
\footnotesize  % Add this to reduce overall font size
\begin{algorithmic}[1]
\STATE \textbf{Input:} EEG segments $\{{\mathbf X}_k\}_{k=1}^K$, speaker features $\{{\mathbf S_1}_k\}_{k=1}^K$ and $\{{\mathbf S_2}_k\}_{k=1}^K$, number of components $Q$
\STATE \textbf{Initialize:} Randomly initialize $\hat{\mathbf W}$
\WHILE{not converged}
\STATE Compute $\tilde\rho_{1k}$ and $\tilde\rho_{2k}$ for each $k$ using \eqref{eq:single_enc_rho} and estimate parameters $\{\mu_a, \sigma_a^2, \mu_u, \sigma_u^2\}$ as in \cite{Lopez2025Unsupervised}.
\STATE Estimate soft labels $p_{1k}, p_{2k}$ based on \eqref{eq:soft_prob_1}-\eqref{eq:soft_prob_2}.
\STATE Build $\mathbf R$, $\mathbf D$ using the soft labels:
\vspace{-0.5em}
\begin{equation*}
  \mathbf R = \begin{bmatrix}
    \mathbf R_{xx} & \mathbf R_{xa} \\
    \transpose{\mathbf R}_{xa} & \mathbf R_{aa}
    \end{bmatrix}, \mathbf D= \begin{bmatrix}
        \mathbf  R_{xx} & \mathbf 0 \\
        \mathbf 0 & \mathbf  R_{aa}
        \end{bmatrix}, 
\end{equation*}
\vspace{-0.5em}
where
\vspace{-0.5em}
\begin{equation*}
    \begin{aligned}
\mat{R}_{xa} &= \sum_{k=1}^K \transpose{\mathbf X}_k(p_{1k} \mathbf S_{1k} + p_{2k}\mathbf S_{2k}),\\
\mat{R}_{aa} &= \sum_{k=1}^K \transpose{(p_{1k}{\mathbf S}_{1k} + p_{2k}{\mathbf S}_{2k} )}(p_{1k}{\mathbf S}_{1k} + p_{2k}{\mathbf S}_{2k}).
    \end{aligned}
\end{equation*}
\vspace{-0.5em}
\STATE Update $\hat{\mathbf W}$ by solving GEVD: $\mathbf R \hat{\mathbf W} = \mathbf D \hat{\mathbf W} \boldsymbol \Lambda$.
\ENDWHILE
\STATE \textbf{Output:} $\hat{\mathbf W}$
\end{algorithmic}
\end{algorithm}

\subsection{Soft Version}
\label{sec:method_soft}

\par Instead of making hard assignments of attended and unattended segments, we propose a soft version that assigns weights to each segment based on prediction uncertainty. This approach provides a principled middle ground between the single-encoder and two-encoder versions, maintaining the more optimal single-encoder structure while incorporating information from both attended and unattended segments adaptively when the model is less certain. The formulation follows \eqref{eq:cca}-\eqref{eq:single_param_gevd}, but replaces $\mathbf S_a$ with its soft version $p_{1k}{\mathbf S}_{1k} + p_{2k}{\mathbf S}_{2k}$, where $p_{1k}$ and $p_{2k}$ are the probabilities of $\mathbf S_{1k}$ and $\mathbf S_{2k}$ being the attended speaker in segment $k$. The statistics $\mat{R}_{xa}$ and $\mat{R}_{aa}$ become:
\begin{equation}
    \label{eq:soft_stats}
    \begin{aligned}
\mat{R}_{xa} &= \sum_{k=1}^K \transpose{\mathbf X}_k(p_{1k} \mathbf S_{1k} + p_{2k}\mathbf S_{2k}),\\
\mat{R}_{aa} &= \sum_{k=1}^K \transpose{(p_{1k}{\mathbf S}_{1k} + p_{2k}{\mathbf S}_{2k} )}(p_{1k}{\mathbf S}_{1k} + p_{2k}{\mathbf S}_{2k}).
    \end{aligned}
\end{equation}

\par The probabilities are estimated in a subject-specific, unsupervised manner using the method proposed by Lopez-Gordo et al. \cite{Lopez2025Unsupervised}. This approach models the (sum of canonical) correlations between EEG and the features of the attended and unattended speakers as two Gaussian distributions: $\mathcal{N}(\mu_a, \sigma_a^2)$ and $\mathcal{N}(\mu_u, \sigma_u^2)$. The parameters of these distributions are estimated from the same training data used to learn the encoder and decoder without knowing the labels (see \cite{Lopez2025Unsupervised} for details). Let $j$ be the index of the attended speaker. We assume no prior label information, i.e., $p(j=1)=p(j=2)=0.5$. Using Bayes theorem, $p_{1k}$ and $p_{2k}$ can be estimated as:
\begin{equation}
\begin{aligned}
    \label{eq:soft_prob_1}
    &p_{1k}= \frac{p(\tilde\rho_{1k},\tilde\rho_{2k}|j=1)p(j=1)}{\sum_{j =\{1,2\}}p(\tilde\rho_{1k},\tilde\rho_{2k}|j)p(j)}=
    \\& \frac{p(\tilde\rho_{1k};\mu_a,\sigma_a^2)p(\tilde\rho_{2k};\mu_u,\sigma_u^2)}{p(\tilde\rho_{1k};\mu_a,\sigma_a^2)p(\tilde\rho_{2k};\mu_u,\sigma_u^2)+p(\tilde\rho_{1k};\mu_u,\sigma_u^2)p(\tilde\rho_{2k};\mu_a,\sigma_a^2)},\\
\end{aligned}
\end{equation}
\begin{equation}
\label{eq:soft_prob_2}
    p_{2k} = 1 - p_{1k},
\end{equation}
where $p(\cdot;\mu,\sigma^2)$ is the probability density function of a Gaussian distribution with mean $\mu$ and variance $\sigma^2$. The soft version is summarized in Algorithm \ref{alg:soft_enc}.

\subsection{Sum-Initialized Single-Encoder}
\label{sec:method_suminit}

\par Rather than randomly initializing the encoder and decoder, we propose training the single-encoder model with a composite signal—the sum of the features of both speakers—in the first iteration. This initialization corresponds to setting $p_{1k}=p_{2k}=0.5$ in the soft version of Section \ref{sec:method_soft}, which allows the model to capture neural responses that are common to both attended and unattended speakers, providing an informative starting point without a bias to a specific speaker for an individual segment $k$. The underlying hypothesis for this simple heuristic is that the bias found in \cite{Heintz2023Unbiased} is mainly driven by the initialization, resulting in a self-sustaining label bias from which the iterations cannot escape.

\begin{figure}[htbp]
    \centering
    \includegraphics[width=\linewidth]{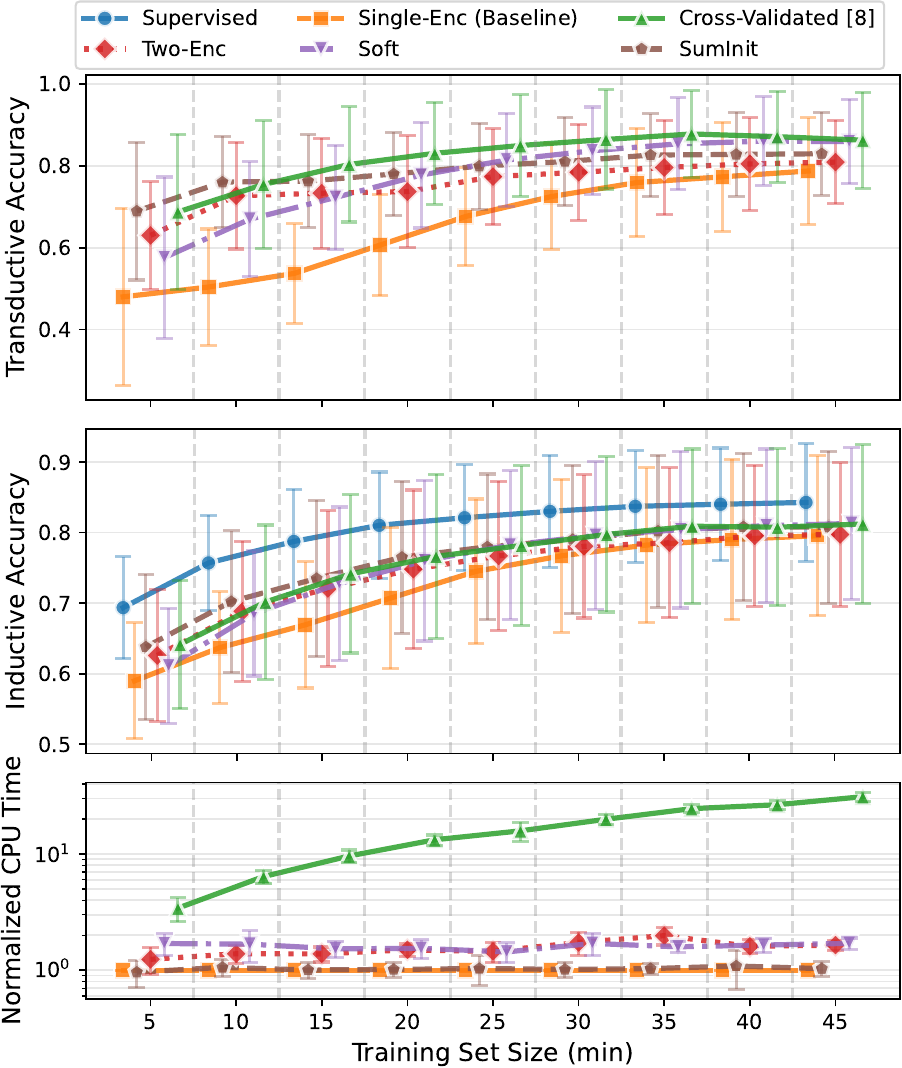}
    \caption{Transductive accuracy, inductive accuracy, and normalized CPU time (w.r.t. baseline) across training set sizes. Dots and bars show mean and standard deviation across subjects and random seeds. Note: the supervised model is not shown in the transductive setting as this would imply using training labels.}
    \label{fig:res}
\end{figure}

\section{Experiments}
\label{sec:experiment}

\subsection{Dataset and Hyperparameters}
\par We evaluate\footnote{For conciseness, we report results for a single dataset. Similar results for an additional dataset are available in the repository linked in the abstract.} all methods on a public dataset from \cite{biesmans2016auditory} used in \cite{Geirnaert2021Unsupervised,Heintz2023Unbiased}. It contains 72-min 64-channel EEG recordings (all channels used) from 16 normal-hearing subjects attending to one of two competing speakers at $\pm 90^\circ$ azimuth, with corresponding audio data. Following \cite{Geirnaert2021Unsupervised}, audio signals are processed using a gammatone filterbank, envelopes are extracted via power-law operation (exponent 0.6) and summed across subbands. Both EEG and speech envelopes are filtered to 1-9 Hz \cite{golumbic2013mechanisms}, downsampled to 20 Hz, and cut into 60-s segments.

\par This paper works with CCA-based models, and thus the hyperparameters are slightly different from \cite{Geirnaert2021Unsupervised}. The number of components $Q$ is set to 2. For EEG signals, we create time-lagged copies at 0-150 ms (capturing current and future information) and stack them along the channel dimension. For audio envelopes, we create time-lagged copies at -250-0 ms (capturing past and current information) and stack them along the feature dimension.

\subsection{Results}

\par We evaluate decoding performance in both transductive and inductive settings. In the transductive setting, predictions are generated for the data on which the unsupervised model is trained, while inductive decoding assesses model generalization to unseen data. We also report normalized CPU time, defined as the ratio of each method's computational time (Windows 11, Intel Core i7-13700F, single thread) to that of the baseline single-encoder approach from Section \ref{sec:method_se}. To examine the performance under limited training data settings, random 3-fold cross-validation is used, with the training sets subsampled to target durations. Train/test splits are identical across methods.

\par As shown in Fig.~\ref{fig:res}, trends are similar for both decoding settings, with the effects of removing initialization bias more pronounced in transductive decoding. The sum-initialized single-encoder consistently outperforms the two-encoder method and is particularly strong with limited data (5-15 min). The soft-label method underperforms on small sets but approaches the cross-validated variant with more data. In computational cost, the normalized CPU time for the cross-validated variant scales linearly with training set size, reaching $\sim 30\times$ for 45-min training sets. All our proposed alternative methods eliminate this scaling: two-encoder and soft methods maintain a constant normalized time of $\sim 1.5\times $ regardless of data size, while the sum-initialized method matches the baseline's time cost ($1.0\times $).

\section{Conclusion}

\par We proposed three computationally efficient solutions to mitigate initialization bias in unsupervised self-adaptive AAD. The two-encoder version trains encoders for both attended and unattended features. The soft version replaces hard segment assignments with probabilistic weights. The sum-initialized single-encoder method initializes the model with a composite signal. For smaller datasets, the sum-initialized approach is the top performer while matching the baseline's computational cost. With larger datasets, the soft-label method becomes competitive, approaching the cross-validated variant's accuracy at low cost.

% A limitation of this work is that evaluation was restricted to a single dataset. Future validation on more diverse datasets is needed to confirm generalizability.

% References should be produced using the bibtex program from suitable
% BiBTeX files (here: strings, refs, manuals). The IEEEbib.bst bibliography
% style file from IEEE produces unsorted bibliography list.
% -------------------------------------------------------------------------
\bibliographystyle{IEEEbib}
\bibliography{strings,refs}

@article{geirnaert2021electroencephalography,
  title={Electroencephalography-based auditory attention decoding: Toward neurosteered hearing devices},
  author={Geirnaert, Simon and Vandecappelle, Servaas and Alickovic, Emina and De Cheveigne, Alain and Lalor, Edmund and Meyer, Bernd T and Miran, Sina and Francart, Tom and Bertrand, Alexander},
  journal={IEEE Signal Processing Magazine},
  volume={38},
  number={4},
  pages={89--102},
  year={2021},
  publisher={IEEE}
}

@article{o2015attentional,
  title={Attentional selection in a cocktail party environment can be decoded from single-trial {EEG}},
  author={O'sullivan, James A and Power, Alan J and Mesgarani, Nima and Rajaram, Siddharth and Foxe, John J and Shinn-Cunningham, Barbara G and Slaney, Malcolm and Shamma, Shihab A and Lalor, Edmund C},
  journal={Cerebral cortex},
  volume={25},
  number={7},
  pages={1697--1706},
  year={2015},
  publisher={Oxford University Press}
}

@article{wong2018comparison,
  title={A comparison of regularization methods in forward and backward models for auditory attention decoding},
  author={Wong, Daniel DE and Fuglsang, S{\o}ren A and Hjortkj{\ae}r, Jens and Ceolini, Enea and Slaney, Malcolm and De Cheveigne, Alain},
  journal={Frontiers in neuroscience},
  volume={12},
  pages={531},
  year={2018},
  publisher={Frontiers Media SA}
}

@article{ciccarelli2019comparison,
  title={Comparison of two-talker attention decoding from {EEG} with nonlinear neural networks and linear methods},
  author={Ciccarelli, Gregory and Nolan, Michael and Perricone, Joseph and Calamia, Paul T and Haro, Stephanie and O’sullivan, James and Mesgarani, Nima and Quatieri, Thomas F and Smalt, Christopher J},
  journal={Scientific reports},
  volume={9},
  number={1},
  pages={11538},
  year={2019},
  publisher={Nature Publishing Group UK London}
}

@article{de2018decoding,
  title={Decoding the auditory brain with canonical component analysis},
  author={De Cheveign{\'e}, Alain and Wong, Daniel DE and Di Liberto, Giovanni M and Hjortkj{\ae}r, Jens and Slaney, Malcolm and Lalor, Edmund},
  journal={NeuroImage},
  volume={172},
  pages={206--216},
  year={2018},
  publisher={Elsevier}
}

@ARTICLE{Geirnaert2021Unsupervised,
  author={Geirnaert, Simon and Francart, Tom and Bertrand, Alexander},
  journal={IEEE Journal of Biomedical and Health Informatics}, 
  title={Unsupervised Self-Adaptive Auditory Attention Decoding}, 
  year={2021},
  volume={25},
  number={10},
  pages={3955-3966},
  doi={10.1109/JBHI.2021.3075631}}

@INPROCEEDINGS{Heintz2023Unbiased,
  author={Heintz, Nicolas and Geirnaert, Simon and Francart, Tom and Bertrand, Alexander},
  booktitle={ICASSP 2023 - 2023 IEEE International Conference on Acoustics, Speech and Signal Processing (ICASSP)}, 
  title={Unbiased Unsupervised Stimulus Reconstruction for {EEG}-Based Auditory Attention Decoding}, 
  year={2023},
  volume={},
  number={},
  pages={1-5},
  doi={10.1109/ICASSP49357.2023.10096608}}

@ARTICLE{Lopez2025Unsupervised,
  author={Lopez-Gordo, Miguel A. and Geirnaert, Simon and Bertrand, Alexander},
  journal={IEEE Transactions on Biomedical Engineering}, 
  title={Unsupervised Accuracy Estimation for Brain-Computer Interfaces Based on Selective Auditory Attention Decoding}, 
  year={2025},
  volume={},
  number={},
  pages={1-12},
  doi={10.1109/TBME.2025.3542253}}

@article{biesmans2016auditory,
  title={Auditory-inspired speech envelope extraction methods for improved {EEG}-based auditory attention detection in a cocktail party scenario},
  author={Biesmans, Wouter and Das, Neetha and Francart, Tom and Bertrand, Alexander},
  journal={IEEE transactions on neural systems and rehabilitation engineering},
  volume={25},
  number={5},
  pages={402--412},
  year={2016},
  publisher={IEEE}
}

@article{geirnaert2022time,
  title={Time-adaptive unsupervised auditory attention decoding using {EEG}-based stimulus reconstruction},
  author={Geirnaert, Simon and Francart, Tom and Bertrand, Alexander},
  journal={IEEE Journal of Biomedical and Health Informatics},
  volume={26},
  number={8},
  pages={3767--3778},
  year={2022},
  publisher={IEEE}
}

@book{corrochano2005handbook,
  title={Handbook of geometric computing: applications in pattern recognition, computer vision, neuralcomputing, and robotics},
  author={Corrochano, Eduardo Bayro},
  year={2005},
  publisher={Springer}
}

@inproceedings{pan2024neuroheed,
  title={NeuroHeed+: Improving neuro-steered speaker extraction with joint auditory attention detection},
  author={Pan, Zexu and Wichern, Gordon and Germain, Fran{\c{c}}ois G and Khurana, Sameer and Le Roux, Jonathan},
  booktitle={ICASSP 2024-2024 IEEE International Conference on Acoustics, Speech and Signal Processing (ICASSP)},
  pages={11456--11460},
  year={2024},
  organization={IEEE}
}

@article{alickovic2019tutorial,
  title={A tutorial on auditory attention identification methods},
  author={Alickovic, Emina and Lunner, Thomas and Gustafsson, Fredrik and Ljung, Lennart},
  journal={Frontiers in neuroscience},
  volume={13},
  pages={153},
  year={2019},
  publisher={Frontiers Media SA}
}

@article{golumbic2013mechanisms,
  title={Mechanisms underlying selective neuronal tracking of attended speech at a “cocktail party”},
  author={Golumbic, Elana M Zion and Ding, Nai and Bickel, Stephan and Lakatos, Peter and Schevon, Catherine A and McKhann, Guy M and Goodman, Robert R and Emerson, Ronald and Mehta, Ashesh D and Simon, Jonathan Z and others},
  journal={Neuron},
  volume={77},
  number={5},
  pages={980--991},
  year={2013},
  publisher={Elsevier}
}

\end{document}